\documentclass[runningheads]{llncs}

\usepackage[utf8]{inputenc}
\usepackage{graphicx}
\usepackage{booktabs}
\usepackage{multirow}
\usepackage{amsmath}
\usepackage{url}

\begin{document}

\title{Domain-Adaptive Dense Retrieval for Brazilian Legal Search}
\titlerunning{Domain-Adaptive Dense Retrieval for Brazilian Legal Search}

\author{Jayr Pereira\inst{1} \and Roberto Lotufo\inst{2} \and Luiz Bonifacio\inst{2}}

\authorrunning{Pereira et al.}

\institute{Universidade Federal do Cariri (UFCA), Juazeiro do Norte, CE, Brazil\\
\email{jayr.pereira@ufca.edu.br}
\and NeuralMind.ai, Campinas, SP, Brazil}

\maketitle

\begin{abstract}

Brazilian legal retrieval is heterogeneous, covering case law, legislation, and question-based search. This makes training dense retrievers a trade-off between stronger domain specialization and broader robustness across retrieval types of search. In this paper, we explore this trade-off using three training setups based on Qwen3-Embedding-4B: a base model with no fine-tuning, a version trained only on legal data, and a mixed setup that combines legal data with SQuAD-pt supervised dataset.
We evaluate these models on five legal datasets from the \textsc{JU\'A} leaderboard, along with \textsc{Quati} dataset as an extra Portuguese retrieval benchmark to test out-of-domain generalization. The legal-only model performs best on the most specialized legal tasks. The mixed setup keeps strong performance on legal data while offering a better overall balance, improving average NDCG@10 from 0.414 to 0.447, MRR@10 from 0.586 to 0.595, and MAP@10 from 0.270 to 0.308 across all six datasets. The biggest improvement appears on \textsc{Quati}, where the mixed model clearly outperforms the legal-only one. Overall, the results show that legal-only and mixed training lead to different strengths: the first is better for specialization, while the second is more robust across different types of search, especially question-based ones. Both adapted models are available on Hugging Face\footnote{Legal Only: \url{https://huggingface.co/ufca-llms/jua-4B-legal-only} and Mixed: \url{https://huggingface.co/ufca-llms/jua-4B-mixed}}.
\end{abstract}

\section{Introduction}

Legal AI systems are increasingly used to support tasks such as case law search, statutory and regulatory lookup, question answering over official legal materials, legal research assistance, and retrieval-augmented generation grounded in authoritative sources \cite{he2026llmlegaltasks,magesh2025hallucinationfree}. In many of these applications, legal information retrieval (LIR) is the first step in the pipeline, so downstream behavior depends directly on the quality of the evidence retrieved at that stage \cite{karpukhin2020dpr,he2026llmlegaltasks,magesh2025hallucinationfree}. At the same time, LIR is not simply a domain-flavored instance of general retrieval. Legal relevance is shaped by authority, procedural context, institutional use, and professional information needs rather than by topical similarity alone \cite{vanopijnen2017relevance}. This makes retrieval in the legal domain an especially demanding setting for dense models, which must capture semantic relatedness while remaining sensitive to highly localized conventions of wording, document structure, and legal function.

These difficulties are magnified in Brazilian Portuguese. Publicly discussed retrieval resources span jurisprudence, legislative proposals, normative acts, and question--answer materials, each with distinct corpora, query styles, and relevance assumptions \cite{juristcu,vitorio,br-tax-qa,jua_preprint}. In practice, this means that legal retrieval in Portuguese is not a single task, but a family of related retrieval regimes. Jurisprudence search often depends on concise institutional summaries; normative retrieval must deal with long, hierarchical documents; and question-driven legal search places greater weight on semantic matching under natural-language formulations \cite{juristcu,vitorio,br-tax-qa}. A retriever calibrated too narrowly to one of these regimes may therefore perform well in-domain while degrading when the corpus structure or query distribution changes.

Prior work in legal retrieval already reflects this tension. Domain-adapted dense encoders can outperform more general-purpose base models in case retrieval \cite{ma2023caseencoder}. Retrieval quality can also improve when the legal structure is explicitly modeled \cite{ma2024structural,louis2023finding}. At the same time, neural approaches for case law and statute law remain sensitive to long documents, query realism, and evaluation design \cite{rossi2021legalsearch,santhosh2024ecthr}. In heterogeneous retrieval more broadly, results from settings such as BEIR suggest that models that look strong under one distribution may not remain strong when evaluation conditions change \cite{thakur2021beir}. Taken together, these findings motivate a more specific question for Brazilian legal retrieval: how should a dense retriever be trained when the target environment itself is heterogeneous?

To address this question, we study a dense retriever for Brazilian legal search under alternative training regimes, using the \textsc{JU\'A} benchmark and related evaluation datasets as the empirical setting \cite{jua_preprint}. We build the comparison on Qwen3-Embedding-4B, a strong open-weight embedding model, and evaluate it under three conditions: an untuned base encoder, a legal-only fine-tuning condition, and a mixed-supervision condition. The mixed recipe combines jurisprudence-oriented supervision from \textsc{JU\'A-Juris} \cite{jua_preprint}, legislative supervision derived from \textsc{Ulysses-RFCorpus} documents outside the evaluation split \cite{vitorio}, and general-domain question--passage supervision from SQuAD-pt. This setup allows us to examine a practical question for heterogeneous Brazilian legal retrieval: whether the most useful dense retriever is the most specialized one or the one that remains more balanced across retrieval regimes.

The contribution of this paper is threefold:
\begin{itemize}
\item We present two Brazilian Portuguese legal dense retrievers based on Qwen3-Embedding-4B: a legal-only model aimed at stronger specialization and a mixed-supervision model aimed at broader cross-regime robustness.
\item We report cross-dataset results over heterogeneous legal retrieval settings and show that the mixed-supervision model largely preserves legal-domain effectiveness while improving substantially on broader and more question-driven retrieval settings.
\item Based on these results, we argue that model selection for legal dense retrieval should be application-oriented: specialized legal workflows may prefer the legal-only profile, whereas heterogeneous search environments benefit from the more robust mixed-supervision profile.
\end{itemize}

The remainder of this paper is organized as follows. Section~\ref{sec:related-work} situates the study within prior work on legal retrieval and dense encoders. Section~\ref{sec:method} presents the research design, training regimes, and evaluation setting. Section~\ref{sec:results} reports the main results and discusses the effect of mixed supervision across datasets. Section~\ref{sec:conclusion} concludes with the main implications and limitations of the study.

\section{Related Work}
\label{sec:related-work}

Research on legal retrieval has consistently emphasized that the domain differs from general retrieval not only in terminology, but also in how relevance itself is shaped by legal authority, procedural context, institutional role, and the practical task faced by the user \cite{vanopijnen2017relevance}. As a result, legal retrieval models do not learn relevance in the abstract: they learn the notion of relevance implicit in their training data and evaluation setting. Survey work on legal case retrieval reinforces this point by highlighting the diversity of corpora, query styles, and relevance notions across legal tasks \cite{feng2024lcrsurvey}.

This heterogeneity also helps explain why legal information retrieval continues to rely on different retrieval paradigms. Lexical retrieval systems, such as BM25, rank documents primarily based on term overlap and weighting schemes \cite{robertson2009bm25}, and they remain strong baselines in many retrieval tasks \cite{thakur2021beir}. In legal search, lexical methods are often effective because exact terminology, statutory references, and recurrent institutional phrasing carry substantial signal. However, lexical methods can be less effective when queries and relevant documents use different wording or when relevance depends on semantic relationships that are not well captured by exact term overlap \cite{karpukhin2020dpr}. This is especially relevant in Brazilian legal retrieval, where available evaluation resources span institutional summaries, normative texts, legislative search, and question-driven formulations \cite{juristcu,vitorio,br-tax-qa,jua_preprint}.

Dense retrieval instead represents queries and documents as continuous embeddings, making it better suited to paraphrase, abstraction, and other forms of semantic mismatch \cite{karpukhin2020dpr}. More broadly, recent embedding models have substantially strengthened this retrieval paradigm, with families such as E5 \cite{wang2022e5} and Qwen3 \cite{qwen3embedding} reporting strong results across heterogeneous evaluation suites such as MTEB \cite{muennighoff2022mteb}. This broader progress is relevant here because it makes strong open-weight encoders viable starting points for domain adaptation.

Recent legal-domain studies then make the trade-off between lexical and dense approaches more concrete. CaseEncoder shows that legal-specific pre-training and knowledge-aware sampling can improve case retrieval over more generic dense base models \cite{ma2023caseencoder}. Related work shows that explicitly modeling legal structure can further improve retrieval when cases are long and internally complex \cite{ma2024structural}. Other studies extend this perspective to statute law, graph-based structural retrieval, and neural search over both case law and statutory corpora \cite{louis2023finding,rossi2021legalsearch}. Recent work on Brazilian legal retrieval likewise points to the importance of hierarchical document organization and retrieval granularity for embedding-based approaches \cite{lima2024multilayered}. Taken together, these studies show that dense retrieval in law benefits from domain-aware signals, but also that the notion of ``domain-aware'' is itself multifaceted, involving legal language, document structure, and task formulation.

The remaining question is how such models behave when evaluation is heterogeneous. In general retrieval, benchmarks such as BEIR and MTEB have shown that retrieval performance can vary sharply across datasets even when models appear strong in aggregate \cite{thakur2021beir,muennighoff2022mteb}. In legal retrieval, this issue is arguably more acute because differences in corpus design and query realism are often more substantial. The ECtHR-PCR benchmark, for example, highlights that precedent retrieval depends on realistic query construction, long document handling, and temporal variation \cite{santhosh2024ecthr}. Transformer-based work in COLIEE-style case retrieval similarly suggests that strong performance depends not only on encoder quality, but also on how retrieval is framed and evaluated \cite{kim2024coliee}. For Portuguese legal retrieval, the same issue appears in a different form: datasets such as \textsc{JurisTCU} \cite{juristcu}, \textsc{Ulysses-RFCorpus} \cite{vitorio}, \textsc{BR-TaxQA-R} \cite{br-tax-qa}, and \textsc{JU\'A} \cite{jua_preprint} make the heterogeneity of the domain empirically visible.

This is the setting examined in the present study. Using Brazilian Portuguese benchmarks spanning different legal retrieval regimes, we compare alternative training conditions for the same dense encoder to analyze how mixed supervision affects the balance between legal specialization and cross-regime robustness.

\section{Method}
\label{sec:method}

\subsection{Research Design}

The central empirical question of this paper is whether the most useful dense retriever for heterogeneous Brazilian legal search is the most specialized one or the most balanced one. To study this question, we compare three training conditions built on the same base encoder, Qwen3-Embedding-4B: an untuned base model, a legal-only fine-tuning condition, and a mixed-supervision condition. This design allows us to compare alternative adaptation regimes while examining whether broader semantic supervision is associated with greater cross-regime robustness.

The comparison is structured to answer two related questions. First, does legal-domain adaptation improve retrieval quality relative to the untuned encoder? Second, once legal adaptation is introduced, does adding a limited amount of general-domain supervision help or hurt performance when the evaluation environment includes multiple legal retrieval regimes? Framed this way, the model variants function as experimentally meaningful conditions for analyzing the relation between specialization and robustness.

\subsection{Training Regimes}

All three conditions use Qwen3-Embedding-4B as the underlying encoder. This base model was chosen for two reasons. First, it belongs to a recent family of open-weight embedding models designed for strong retrieval and reranking performance, with competitive results on broad evaluation suites such as MTEB \cite{muennighoff2022mteb,qwen3embedding}. Second, in the \textsc{JU\'A} evaluation setting, it provides a strong open baseline, outperforming other open-weight dense retrievers such as KaLM Gemma3 12B \cite{jua_preprint}. The base condition corresponds to the untuned model. The legal-only condition corresponds to a fine-tuning run primarily supervised by the legal domain, without the additional SQuAD-based component used in the mixed recipe. The mixed condition corresponds to a mixed-supervision model intended for heterogeneous retrieval settings where legal specialization and broader semantic robustness are both desirable. 

The mixed training regime combines three main supervision sources: \textsc{JU\'A-Juris} train, Ulysses-derived legislative supervision, and SQuAD-pt. The Ulysses portion also includes a small synthetic extension based on alternative automatically generated query formulations from the same legislative collection. The goal of this mixture is to expose the encoder to different forms of relevance rather than to a single legal distribution. \textsc{JU\'A-Juris} provides jurisprudence-oriented supervision in which short legal statements are paired with supporting passages from judicial decisions, approximating case-law retrieval based on concise institutional summaries. Ulysses train contributes legislative retrieval pairs built from bills that are not used in the evaluation split: the legislative summary (\textit{ementa}) is used as the query, and the full bill text is treated as the positive document \cite{vitorio}. The small synthetic extension increases coverage while preserving the same legislative retrieval setting. SQuAD-pt introduces question--passage supervision from a broader domain, contributing more varied natural-language formulations and a less institutionally constrained query distribution.

SQuAD-pt was also chosen partly for reasons of scale. A substantially larger alternative, such as mMARCO-pt \cite{bonifacio2021mmarco}, would have dominated the legal datasets available in this study unless an additional sampling or selection stage were introduced. That, in turn, would have added a further source of methodological bias to the experimental design. In this sense, SQuAD-pt provided a smaller and more controlled source of Portuguese question--passage supervision.

The Ulysses training portions are therefore constructed from legislative bills that do not overlap with the relevance-annotated split used for evaluation. For each remaining bill, the legislative summary (\textit{ementa}) is used as the basis for query creation, while the full bill text is treated as the positive passage. In addition to the original summary-based query, we generate a synthetic query variant from the same \textit{ementa}, producing alternative formulations over the same underlying relevance relation. We then apply an additional filtering step to the Ulysses and Ulysses synthetic portions, retaining only examples whose positive document is effectively recovered by a first-stage BM25 run and prioritizing instances with better rank, stronger score margin, and richer negative pools. Besides improving training quality, this step removes many redundant or overly similar query formulations in practice. The resulting balanced splits contain 42,580 Ulysses examples and 2,101 Ulysses synthetic examples.

This distinction is important for interpreting the results reported later. In the empirical comparison, the legal-only condition is slightly stronger on some of the most specialized subsets, whereas the mixed condition is more robust in the aggregate and substantially stronger on broader retrieval settings such as \textsc{Quati}. In this sense, SQuAD-pt is not treated as a generic source of additional data, but as a regularizing component in the training design.

\subsection{Training Procedure}

Fine-tuning is performed with \texttt{ms-swift} \cite{zhao2024swift} using LoRA \cite{hu2021lora} and an InfoNCE-style contrastive objective \cite{oord2018cpc}. Because this objective benefits from hard negatives semantically competitive with the positive passage, we construct training instances from first-stage BM25 retrieval runs, combining in-batch negatives with explicit hard negatives. For \textsc{JU\'A-Juris} and SQuAD-pt, the hard-negative construction follows the same general pattern. For each query, a BM25 run retrieves a ranked candidate list, and negatives are selected from the top retrieved documents after removing the positive passage. 

In both cases, the selection is based on a statistical cutoff over BM25 scores, so that negatives are concentrated on documents that remain competitive under a sparse first-stage retriever. In SQuAD-pt, we additionally discard very short or ambiguous questions before building the training set, which helps avoid noisy supervision from underspecified queries. The final mixed dataset contains 89,362 training instances: 27,690 from \textsc{JU\'A-Juris}, 42,580 from Ulysses train, 2,101 from Ulysses synthetic train, and 16,991 from SQuAD-pt. In practical terms, the resulting mixture is dominated by legal supervision, especially jurisprudential and legislative retrieval, while retaining a smaller but non-trivial amount of broader question--passage supervision.

\section{Evaluation Protocol}

\subsection{Datasets}

The evaluation is designed to reflect the heterogeneity discussed in the previous sections. We therefore consider the five legal datasets that make up the \textsc{JU\'A} leaderboard environment, together with \textsc{Quati} as an additional Portuguese retrieval benchmark for out-of-domain generalization:
\begin{itemize}
\item \textbf{\textsc{JU\'A-Juris}}: jurisprudence retrieval over curated TCU jurisprudence excerpts. Queries are \textit{enunciados}, that is, abstractive summaries of rulings, and relevance follows a binary protocol that pairs each summary with its supporting excerpt \cite{jua_preprint}.
\item \textbf{\textsc{JurisTCU}}: jurisprudence retrieval over TCU case-law excerpts with expert-verified relevance judgments. The dataset includes real keyword-style queries together with synthetic variants, making it a second but distinct jurisprudence-oriented regime \cite{juristcu,jua_preprint}.
\item \textbf{\textsc{NormasTCU}}: retrieval over TCU normative acts. It represents a regulatory retrieval setting with long, hierarchical documents, short keyword queries, richer formulations, and three-level graded relevance judgments \cite{jua_preprint}.
\item \textbf{\textsc{Ulysses-RFCorpus}}: legislative retrieval benchmark based on real relevance feedback from the Brazilian Chamber of Deputies. It uses user-oriented legislative queries over long parliamentary documents, representing a more institutionally grounded legislative retrieval regime \cite{vitorio,jua_preprint}.
\item \textbf{\textsc{BR-TaxQA-R}}: question-driven tax retrieval over tax answers and linked reference material. Its FAQ-style questions and graded relevance judgments make it the clearest legal QA-like retrieval setting in the benchmark \cite{br-tax-qa,jua_preprint}.
\item \textbf{\textsc{Quati}}: a general-domain Brazilian Portuguese retrieval benchmark whose queries were written by native speakers and whose corpus was curated from frequently accessed Brazilian websites. We use it as an additional non-legal reference point for out-of-domain generalization \cite{bueno}.
\end{itemize}

This combination allows us to examine not only performance within legal search, but also whether a domain-adapted retriever remains robust when query style and corpus structure change. We report Normalized Discounted Cumulative Gain (NDCG@10), Mean Reciprocal Rank (MRR@10), and Mean Average Precision (MAP@10), where `@10' indicates that evaluation is truncated to the top 10 retrieved results. NDCG@10 emphasizes the quality of the ranking near the top of the list, MRR@10 captures how early the first relevant result appears, and MAP@10 reflects the overall precision profile of the ranked list within the first 10 positions.

\section{Results}
\label{sec:results}

\subsection{Main Findings}

Table~\ref{tab:main_results} reports the core comparison in this paper: the untuned base encoder, the legal-only adaptation condition, and the mixed-supervision condition. Since all three are available on the same five legal datasets from the \textsc{JU\'A} leaderboard and on \textsc{Quati}, the comparison can be carried out on a shared six-dataset setting using the metrics described above.

\begin{table*}[t]
\centering
\scriptsize
\setlength{\tabcolsep}{3pt}
\resizebox{\textwidth}{!}{%
\begin{tabular}{lccccccccc}
\toprule
\multirow{2}{*}{Dataset} & \multicolumn{3}{c}{NDCG@10} & \multicolumn{3}{c}{MRR@10} & \multicolumn{3}{c}{MAP@10} \\
\cmidrule(lr){2-4}\cmidrule(lr){5-7}\cmidrule(lr){8-10}
 & Base & Legal & Full & Base & Legal & Full & Base & Legal & Full \\
\midrule
\textsc{JU\'A-Juris} & 0.199 & \textbf{0.294} & 0.290 & 0.152 & \textbf{0.233} & 0.230 & 0.152 & \textbf{0.233} & 0.231 \\
\textsc{JurisTCU} & 0.311 & \textbf{0.375} & 0.363 & 0.588 & \textbf{0.650} & 0.641 & 0.138 & \textbf{0.179} & 0.170 \\
\textsc{NormasTCU} & 0.307 & \textbf{0.310} & 0.305 & \textbf{0.508} & 0.461 & 0.474 & \textbf{0.199} & 0.186 & 0.184 \\
\textsc{Ulysses} & \textbf{0.450} & 0.426 & 0.441 & \textbf{0.719} & 0.619 & 0.624 & 0.233 & 0.301 & \textbf{0.315} \\
\textsc{BR-TaxQA} & 0.771 & 0.756 & \textbf{0.777} & 0.797 & 0.779 & \textbf{0.800} & 0.693 & 0.677 & \textbf{0.701} \\
\textsc{Quati} & 0.447 & 0.438 & \textbf{0.503} & 0.754 & 0.770 & \textbf{0.799} & 0.205 & 0.197 & \textbf{0.247} \\
\midrule
Average & 0.414 & 0.433 & \textbf{0.447} & 0.586 & 0.585 & \textbf{0.595} & 0.270 & 0.296 & \textbf{0.308} \\
\bottomrule
\end{tabular}
}
\caption{Results for the three conditions emphasized in this paper. `Base' denotes Qwen/Qwen3-Embedding-4B, `Legal' denotes the legal-only adaptation condition, and `Full' denotes the released mixed-supervision model. Boldface marks the best value within each row and metric block.}
\label{tab:main_results}
\end{table*}

The results support three main findings. First, legal-domain supervision clearly improves retrieval relative to the untuned encoder. Both adapted conditions outperform the base model on the average NDCG@10 comparison, and the gains are especially visible on \textsc{JU\'A-Juris} and \textsc{JurisTCU}. This indicates that adaptation to Brazilian legal material is beneficial even before considering the difference between the legal-only and full training regimes.

Second, the mixed-supervision condition largely preserves legal-domain effectiveness while improving broader-domain robustness. Relative to the legal-only condition, the mixed model remains very close on the most specialized legal subsets: it moves from 0.294 to 0.290 on \textsc{JU\'A-Juris}, from 0.375 to 0.363 on \textsc{JurisTCU}, and from 0.310 to 0.305 on \textsc{NormasTCU}. At the same time, it improves over the legal-only condition on \textsc{Ulysses-RFCorpus}, \textsc{BR-TaxQA-R}, and especially \textsc{Quati}, where NDCG@10 rises from 0.438 to 0.503, MRR@10 from 0.770 to 0.799, and MAP@10 from 0.197 to 0.247. The overall pattern is therefore not one of a large legal-domain sacrifice in exchange for broader generalization, but rather one of modest in-domain differences combined with a substantial gain in a broader retrieval regime.

Third, adding SQuAD appears to improve robustness without displacing the model's legal specialization. The average gain from 0.433 to 0.447 in NDCG@10 suggests that the mixed regime is preferable when the goal is a reusable retriever rather than a narrowly specialized encoder for jurisprudence retrieval.

At the same time, adaptation is not uniformly beneficial across all legal datasets. On \textsc{NormasTCU}, both adapted conditions underperform the untuned encoder on MRR@10 and MAP@10, and the mixed condition is also slightly below the base model on NDCG@10. On \textsc{Ulysses-RFCorpus}, the untuned encoder remains stronger than both adapted conditions on NDCG@10 and MRR@10, although the mixed model improves MAP@10.

One plausible explanation is that these datasets reward a different retrieval bias from the one strengthened by the mixed training regime. \textsc{NormasTCU} is centered on normative documents whose structure is closer to statutes and regulations than to the jurisprudential materials that dominate the training mixture. It also contains relatively few queries, which makes exact matching behavior more consequential, and BM25 is already a strong baseline on this task in the \textsc{JU\'A} leaderboard environment \cite{jua_preprint}. \textsc{Ulysses-RFCorpus} shows a related, though milder, pattern over long, institutionally structured legislative documents. By contrast, the large gain on \textsc{Quati} suggests that the mixed regime strengthens broader semantic matching under more variable query formulations. On this reading, adding SQuAD-pt does not simply make the retriever ``better'' or ``worse'' in the abstract; it shifts the model toward a better balance between legal specialization and semantic robustness while keeping the legal-domain losses relatively limited.

\subsection{Interpreting the Trade-off}

The contrast between the legal-only and mixed conditions is the clearest evidence for the role of general-domain supervision. If SQuAD-pt were merely additional, unrelated data, the expected outcome would be a pronounced dilution of legal performance with no clear compensating benefit. That is not the pattern observed in the results. Instead, the legal-only condition retains a slight advantage in the most specialized legal subsets, while the mixed condition remains close on those datasets and improves performance on broader retrieval settings. This is more consistent with interpreting SQuAD-pt as a regularizing source of semantic variation.

The behavior of NDCG@10, MRR@10, and MAP@10 reinforces this interpretation. On the more specialized legal subsets, the legal-only condition often improves both NDCG@10 and MRR@10, suggesting that concentrated legal supervision helps move the most relevant item earlier in the ranking. On \textsc{Quati} and on the broader legal tasks, however, the mixed model improves not only NDCG@10 and MRR@10 but also MAP@10, suggesting that the gains are not limited to the earliest relevant hit. In other words, retrieval quality appears to improve under more varied query styles while legal-domain effectiveness is largely preserved.

This trade-off is also visible at the level of the retrieval regime. \textsc{JU\'A-Juris} and \textsc{JurisTCU} are jurisprudence-heavy tasks, where concise institutional summaries and supporting passages make legal phrasing especially important. \textsc{NormasTCU} and \textsc{Ulysses-RFCorpus} represent more structured legal retrieval over normative or legislative materials. \textsc{Quati} \cite{bueno}, and to a lesser extent \textsc{BR-TaxQA-R}, are the datasets in our evaluation that most clearly emphasize broader semantic matching, since they rely more heavily on natural-language question formulations than the jurisprudential and normative collections. Read together, these results suggest that the complete model is the best compromise because it preserves most of the gains from legal specialization while transferring better across regimes.

For downstream applications, this distinction matters because retrieval quality is not exhausted by whether one relevant item appears at rank 1. Systems that operate upstream of reranking or RAG benefit from candidate sets that remain semantically appropriate under different query distributions \cite{karpukhin2020dpr,magesh2025hallucinationfree}. From that perspective, the mixed-supervision model is the stronger choice whenever the retriever is intended to support heterogeneous legal search rather than a single narrowly specialized jurisprudential workflow.

\subsection{Comparison with the Full Leaderboard}

The broader leaderboard results help position the proposed models relative to a wider set of baselines. To make this comparison fair across all models considered there, Table~\ref{tab:backbone_scale} reports averages over the four legal datasets shared by the leaderboard entries: \textsc{JU\'A-Juris}, \textsc{JurisTCU}, \textsc{NormasTCU}, and \textsc{BR-TaxQA-R}.

\begin{table}[t]
\centering
\small
\setlength{\tabcolsep}{4pt}
\begin{tabular}{lccc}
\toprule
Model & NDCG@10 & MRR@10 & MAP@10 \\
\midrule
BM25/anserini & 0.350 & 0.444 & 0.243 \\
text-embedding-3-small & 0.329 & 0.429 & 0.245 \\
KaLM Gemma3 12B & 0.298 & 0.383 & 0.224 \\
Qwen3-Embedding-4B (base) & 0.397 & 0.511 & 0.295 \\
Qwen3-Embedding-8B (base) & 0.407 & 0.509 & 0.304 \\
Qwen3-Embedding-4B (legal-only adapted) & 0.434 & 0.531 & 0.319 \\
\textbf{Qwen3-Embedding-4B (mixed adapted)} & \textbf{0.434} & \textbf{0.536} & \textbf{0.321} \\
\bottomrule
\end{tabular}
\caption{Average performance on the four legal datasets shared by all baselines considered in this comparison.}
\label{tab:backbone_scale}
\end{table}

On this shared comparison, the two adapted Qwen3-Embedding-4B variants outperform the untuned Qwen3-Embedding-8B model. The legal-only adapted model reaches average scores of 0.434 NDCG@10, 0.531 MRR@10, and 0.319 MAP@10, while the mixed adapted model reaches 0.434, 0.536, and 0.321. By comparison, the untuned Qwen3-Embedding-8B model reaches 0.407, 0.509, and 0.304. This indicates that legal adaptation and supervision design can outweigh a simple increase in base model size.

The same pattern appears when the adapted 4B model is compared with other strong baselines available in the leaderboard. It remains above BM25/anserini, text-embedding-3-small, and KaLM Gemma3 12B on all three metrics in the shared legal comparison. Taken together, these results suggest that the main gains reported in this paper are not explained solely by starting from a strong base model. They depend more specifically on how that base model is adapted to the target retrieval environment.

\subsection{Qualitative Examples}

The aggregate metrics can be complemented with a few illustrative queries. In a jurisprudential query from \textsc{JU\'A-Juris} concerning the effect of Article 5 of Law 9.717/1998 on pension eligibility (roughly, ``Did Article 5 of Law 9.717/1998 remove certain statutory civil pension categories from the federal public servants' pension regime?''), the legal-only condition moves the relevant decision from rank 2 to rank 1 relative to the untuned encoder. The top-ranked results in both conditions are legally related, but the legal-only adaptation is more successful in prioritizing the exact precedent rather than a nearby decision with overlapping statutory language. This is consistent with the view that concentrated legal supervision sharpens distinctions that matter in jurisprudence retrieval.

A contrasting pattern appears in \textsc{NormasTCU}. For the query ``principais diretrizes de normas de auditoria do operacional do tribunal de contas da união'' (``main guidelines of operational audit norms in the Federal Court of Accounts''), the untuned encoder places the relevant document at rank 2, whereas the legal-only condition moves it down to rank 10. In this case, the adapted model appears to over-prioritize nearby normative documents with highly similar titles and institutional framing, including other manuals and general norms on operational auditing. This example is consistent with the regressions observed for \textsc{NormasTCU}: adaptation can increase sensitivity to legal language while making fine distinctions among closely related normative texts more difficult.

The opposite behavior appears in broader semantic retrieval. For the \textsc{Quati} query ``Por que dividir um país em estados?'' (``Why divide a country into states?''), the mixed-supervision condition places a relevant document at rank 1, whereas the base and legal-only conditions retrieve the first relevant result only at rank 2. Here, the top-ranked document under the mixed condition directly explains the administrative and political rationale for dividing a country into states, while the other conditions place more weakly aligned educational material ahead of it. This example supports the interpretation that mixed supervision improves retrieval when success depends less on stable institutional phrasing and more on semantic matching across varied formulations.

\subsection{Limitations}

The results should be interpreted in light of four main limitations. First, the contrast between the legal-only and mixed conditions supports an informative ablation, but it should not be read as a fully controlled causal decomposition of every training component. The legal-only condition serves here as the closest available approximation to a purely legal supervision regime. A more exhaustive account of intermediate checkpoints and training variants would allow a finer-grained analysis of which parts of the mixed recipe drive the observed gains.

Second, although the qualitative examples provide some query-level grounding, the evidence reported in this paper remains primarily benchmark-level. This is sufficient to support the main claim that stronger specialization and stronger robustness do not coincide on the same condition, but it does not yet explain systematically which query types, legal formulations, or document structures are most responsible for the observed differences. Broader query-level analysis and qualitative error inspection would therefore be valuable complements to the present results.

Third, a further limitation concerns training scale. Although the mixed training set is heterogeneous, it remains relatively small for adapting a 4B-parameter embedding model, especially when compared with the scale typically used in modern retrieval training. The reported results should therefore be interpreted as evidence about the behavior of different supervision regimes under limited-data adaptation, rather than as an estimate of the best achievable performance for Brazilian legal retrieval.

Fourth, the external validity of the conclusions is bounded by the six evaluation datasets considered here. Although these datasets cover jurisprudential, normative, legislative, question-driven, and broader semantic retrieval regimes, they do not exhaust the range of tasks encountered in deployed legal search systems.

\section{Conclusion}
\label{sec:conclusion}

This paper examined how alternative training conditions affect the behavior of a dense retriever for heterogeneous Brazilian legal search. To do so, we compared an untuned encoder, a legal-only adaptation condition, and a mixed-supervision condition on the five legal datasets in the \textsc{JU\'A} evaluation environment together with \textsc{Quati}. The main result is that the two adapted models occupy different useful points in the specialization--robustness space. The legal-only condition is slightly stronger on some of the most specialized legal subsets, whereas adapting Qwen3-Embedding-4B with mixed legal and general-domain supervision yields the most balanced condition among the three compared here, improving average NDCG@10 from 0.414 for the untuned encoder and 0.433 for the legal-only condition to 0.447 for the mixed-supervision condition over six shared datasets.

The comparison also clarifies the role of SQuAD-pt in the training mixture. Removing it slightly favors the most specialized legal subsets, but the mixed-supervision condition remains close on those datasets while performing better on broader retrieval settings and substantially better on \textsc{Quati}. These findings reinforce that heterogeneous legal retrieval should not be treated as a single optimization target. Taken together, the results support releasing both adapted models: the legal-only model as a specialized option for more institutionally framed legal retrieval, and the mixed model as a more robust option for heterogeneous and question-driven search. Both models are available online. More broadly, the results suggest that model selection in legal retrieval should not be framed only in terms of peak in-domain performance, but also in terms of robustness across retrieval regimes. Two natural directions for future work are to expand the training mixture with additional synthetic legal data and study how that changes the specialization--robustness balance, and to evaluate these retrievers in downstream settings such as RAG and other legal tasks built on top of retrieval.

\bibliographystyle{splncs04}
\bibliography{references}

\end{document}